\documentclass[12pt,sort&compress]{iopart}

\usepackage{iopams}
\usepackage{graphicx}
\usepackage{cite}
\begin{document}

\title[Reentrant phenomenon and inverse magnetocaloric effect \ldots]
{Reentrant phenomenon and inverse magnetocaloric effect
 in a generalized spin-$(1/2,s)$ Fisher's super-exchange antiferromagnet}

\author{Lucia G\'alisov\'a}

\address{Department of Applied Mathematics and Informatics,
		 Faculty of Mechanical Engineering, Technical University of Ko\v{s}ice,
		 Letn\'a 9, 042 00 Ko\v{s}ice, Slovakia}
\ead{galisova.lucia@gmail.com}
\vspace{10pt}
\begin{indented}
\item[]
\end{indented}

\begin{abstract}
The thermodynamic and magnetocaloric properties of a generalized spin-$(1/2,s)$  Fisher's super-exchange antiferromagnet are exactly investigated by using the decoration-iteration mapping transformation. Besides the critical temperature, sublattice magnetization, total magnetization, entropy and specific heat, the isothermal entropy change and adiabatic temperature change are rigorously calculated in order to examine cooling efficiency of the model in a vicinity of the first- and second-order phase transitions. It is shown that an enhanced inverse magnetocaloric effect occurs around the temperature interval $T_c(B\neq 0) \lesssim T < T_c(B = 0)$ for any magnetic-field change $\Delta B: 0 \to B$. The most pronounced inverse magnetocaloric effect can be found nearby the critical field, which corresponds to the zero-temperature phase transition from the long-range ordered ground state to the paramagnetic one. The observed phenomenon increases with the increasing value of decorating spins. Furthermore, sufficiently high values of decorating spins have also linked to a possibility of observing reentrant phase transitions at finite temperatures.
\end{abstract}

%
\vspace{2pc}

%
%
\maketitle

\section{Introduction}
\label{sec:1}

The magnetocaloric effect (MCE), which is defined as the temperature change (i.e., as the cooling or heating) of a magnetic system due to the variation of an external magnetic field, has a long history in cooling applications at various temperature regimes~\cite{War81}. Since the first successful experiment of  the  adiabatic  demagnetization  performed  in 1933~\cite{Gia33}, the MCE represents the standard technique for achieving the extremely low temperatures~\cite{Str07}. In this regard, theoretical predictions and descriptions of materials showing an enhanced MCE create real opportunities for the effective selection of the construction for working magnetic-refrigeration devices. A great theoretical interest in~MCE has recently been focused on some frustrated structures that may achieve huge adiabatic cooling rates in the vicinity of critical fields due to the large (often macroscopic) degeneracy of states~\cite{Zhi03, Zhi04, Der06, Sch07, Hon09a}.

In general, the MCE is characterized by the isothermal entropy change ($\Delta S_T$) and by the adiabatic temperature change~($\Delta T_{ad}$) upon the magnetic field variation. Depending on sings of these magnetocaloric potentials, the MCE can be either conventional ($\Delta S_T<0$, $\Delta T_{ad}>0$) or inverse ($\Delta S_T>0$, $\Delta T_{ad}<0$). In the former case the system cools down when the magnetic field is removed adiabatically, while in the latter case it heats up. Whether the conventional or inverse MCE is present basically depends on the particular magnetic arrangement in the system. Namely, the former phenomenon can be observed in regular ferromagnets or paramagnets, while the latter one can be detected in ferrimagnetic and antiferromagnetic materials. Moreover, the coexistence of both phenomena is also possible. In fact, the conventional and inverse MCEs have been theoretically observed in magnetic systems with rich structure of the ground-state phase diagram, in particular, in various one-dimensional spin models~\cite{Zhi04, Der06, Hon09, Can09, Tri10, Lan10, Hon11, Top12, Ver13, Kas13, Gal14, Str14a, Zar15, Gal15a, Per09, Gal15b, Gal15c, Gal15d}, some finite structures~\cite{Sch07, Hon09a, Str14b, Sha14, Str15a} and multilayers~\cite{Sza14}.
However, the enhanced MCE has been so far rigorously investigated only in one-dimensional systems~\cite{Zhi04, Der06, Hon09, Can09, Tri10, Lan10, Hon11, Top12, Ver13, Kas13, Gal14, Str14a, Zar15, Gal15a, Per09, Gal15b, Gal15c, Gal15d} or in some finite structures~\cite{Str14b,Str15a} due to a lack of exactly solved spin models in higher dimensions accounting for a non-zero magnetic field. Theoretical description of this phenomenon in two- and three-dimensional magnetic systems is usually based only on some approximative methods~\cite{Hu08, Oli08, Ran10, Sza11, Sza14}.

In 1960, M.E. Fisher has proposed a novel spin-$1/2$ super-exchange Ising antiferromagnet on a decorated square lattice, which permits a rigorous solution of the partition function in the presence of an external magnetic field~\cite{Fis60a, Fis60b}. The spin-$1/2$ Fisher's super-exchange model and its other variants~\cite{Hat68, Mas73, Lu05, Can06} can thus be used for exact theoretical study of the effect of applied field on magnetic properties of a certain class of magnetic insulators, e.g., for investigation of the cooling or heating efficiency of the system in a vicinity of discontinuous (first-order) and continuous (second-order) phase transitions. In addition, these spin models may also bring a considerable insight into deficiencies of some approximative methods. Motivated by aforementioned facts, the purpose of this paper is to extend the rigorous theoretical examination of the MCE to a class of two-dimensional spin models. We will consider the generalized spin-$(1/2,s)$ Fisher's super-exchange model on a decorated square lattice in order to bring a deeper insight into how the critical behavior of the model depends on the magnitude of decorating spins. Particular attention will be paid to the examination of regions showing an enhanced~MCE.

The outline of the paper is as follows: In Sec.~\ref{sec:2}, the generalization of the Fisher's super-exchange model together with a brief description of its exact analytical treatment will be carried. In Sec.~\ref{sec:3}, the most interesting results for the ground state, the finite-temperature phase diagram as well as the magnetization, specific heat and entropy will be discussed. Magnetocaloric properties of the model will be presented in detail in Sec.~\ref{sec:4}. Finally, Sec.~\ref{sec:5} will bring some conclusions and future outlooks.

\begin{figure}[t!]
\centering
  \includegraphics[width=0.55\textwidth]{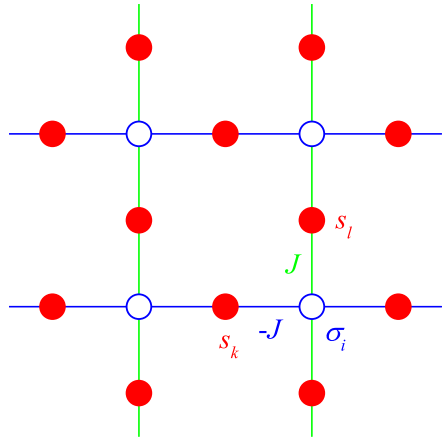}
\vspace{-5mm}
\caption{The spin-$(1/2, s)$ Fisher's super-exchange model on a~decorated square lattice. The empty circles denote nodal lattice sites occupied by the Ising spins $\sigma = 1/2$, while the full circles mark lattice positions of the decorating Ising spins of arbitrary magnitude $s$.}
\label{fig:1}
\end{figure}

\section{Model and its exact solution}
\label{sec:2}

Let us consider a mixed spin-$(1/2, s)$ Ising model on a decorated square lattice involving the effect of an external magnetic field, as is schematically depicted in Fig.~\ref{fig:1}. In this figure, the empty circles denote nodal lattice sites occupied by the Ising spins $\sigma = 1/2$, while the full ones label decorating lattice sites occupied by the Ising spins of an arbitrary magnitude $s$. Because of the two-dimensional Ising model whose all spins are placed into the external magnetic field still represents an unsolvable problem of statistical mechanics, we will further assume the simplified version of the model, in which the longitudinal magnetic field $B$ acts just on decorating spins. In addition, exchange interactions of the same intensity but of opposite signs have to be supposed between the nearest spin neighbors in the horizontal and vertical directions to ensure an exact tractability of the considered spin system. For this reason, we will further assume the ferromagnetic (antiferromagnetic) coupling $-J<0$ ($J>0$) on the horizontal (vertical) bonds of the lattice. Under the above assumptions, the total Hamiltonian of the model reads
\begin{equation}
\label{eq:H}
{\cal H} = -J\sum_{\left\langle i, k\right\rangle}^{N}\sigma_i s_k +  J\sum_{\left\langle i, l\right\rangle}^{N}\sigma_i s_l - B\sum_{k=1}^{N}s_k - B\sum_{l=1}^{N}s_l,
\end{equation}
where $s_{k(l)} = -s, -s+1, \ldots, s$ labels the decorating Ising spin at $k$th horizontal ($l$th vertical) bond and $\sigma_i=\pm1/2$ denotes the nodal Ising spin at $i$th site of the original square lattice. The first (second) summation in the Hamiltonian~(\ref{eq:H}) is carried out over nearest-neighboring lattice sites in the horizontal (vertical) direction, while other two terms represent Zeeman's energies of the decorating spins. Finally, $N$ represents the total number of nodal sites of the original lattice, i.e., the Ising spins~$\sigma$ (we consider the thermodynamic limit $N\to\infty$). It is worth emphasizing that the considered spin model is generally slightly different from the usual antiferromagnetic Ising square lattice. In particular, the standard antiferromagnetic model becomes ferromagnetic when the sign of exchange integral $J$ is changed. By contrast, the Hamiltonian~(\ref{eq:H}) remains invariant against the transformation $J \to -J$.

The two-dimensional spin model defined in the above way is exactly solvable within the framework of a generalized decoration-iteration mapping transformation~\cite{Fis59, Syo72, Str10}. More specifically, different signs of the exchange constants on the horizontal and vertical bonds of the decorated square lattice cancel out contributions of the mapping terms that represent effective magnetic fields acting on nodal spins of the corresponding simple lattice (for more computational details see Fisher's original works~\cite{Fis60a, Fis60b} and our previous work~\cite{Can06}). As a result, one obtains a simple relation between the partition function ${\cal Z}_{F}$ of the considered mixed spin-$(1/2,s)$ Fisher's super-exchange model~(\ref{eq:H}) and the partition function ${\cal Z}_{I}$ of the spin-$1/2$ Ising model on a simple square lattice defined by the Hamiltonian ${\cal H}_{I} = -J_{eff}\sum_{\langle i,j\rangle}^{2N}\sigma_i\sigma_j$\,:
\begin{eqnarray}
\label{eq:Z}
{\cal Z}_{F}(\beta, J, B, s) = A^{2N}{\cal Z}_{I}(\beta, J_{eff}).
\end{eqnarray}
Above, $\beta = 1/(k_{\rm B}T)$ is the inverse temperature ($k_{\rm B}$ is the Boltzmann's constant) and the mapping parameters $A$, $J_{eff}$ are unambiguously determined by a 'self-consistency' condition of the applied decoration-iteration transformation~\cite{note1}.

At this stage, the exact treatment of the generalized Fisher's super-exchange model is formally completed, because the partition function ${\cal Z}_{I}$ of the spin-$1/2$ Ising model on the square lattice is well known~\cite{Ons44}:
\begin{eqnarray}
\label{eq:F_I}
\ln {\cal Z}_{I}  = N\ln 2 + \frac{N}{2\pi^2}\int_{0}^{\pi}\!\int_{0}^{\pi}\ln\left({\cal C}^2 - {\cal S}\cos\theta - {\cal S}\cos\phi\right){\rm d}\theta{\rm d}\phi.
\end{eqnarray}
Here, ${\cal C} = \cosh(\beta J_{eff}/2)$ and ${\cal S} = \sinh(\beta J_{eff}/2)$. Actually, Eq.~(\ref{eq:Z}) in combination with exact mapping theorems developed by Barry {\it et al.}~\cite{Bar88, Kha90, Bar91, Bar95} and the generalized Callen-Suzuki spin identity~\cite{Cal63, Suz65, Bal02} allow us to rigorously express the spontaneous magnetization $m_\sigma$ of the nodal spins, as well as the magnetization $m_{h}$, $m_{v}$ of the decorating spins located on horizontal and vertical bonds, respectively:
\begin{eqnarray}
\label{eq:m_sigma}
m_{\sigma} &\equiv \langle\sigma_i\rangle  = \langle\sigma_i\rangle_I \equiv m_I,
\\
\label{eq:m_sh}
m_{h} &\equiv \langle s_k\rangle  = K - 4m_I L + 4c_I M,
\\
\label{eq:m_sv}
m_{v} &\equiv \langle s_l\rangle  = K + 4m_I L + 4c_I M.
\end{eqnarray}
Above, the symbols $\langle\cdots\rangle$ and $\langle\cdots\rangle_I$ denote the standard canonical ensemble average performed over the generalized spin-$(1/2,s)$ Fisher's model~(\ref{eq:H}) and the corresponding spin-$1/2$ Ising model on a square lattice, respectively. Obviously, the aforelisted magnetization are expressed in terms of the spontaneous magnetization $m_I$ and the two-spin correlation function $c_I$ between nearest-neighboring spins of the spin-$1/2$ Ising square lattice. Since rigorous solutions for both quantities are well known~\cite{Kau49, Yan52}, we can restrict ourselves just for appointment of the coefficients $K$, $L$, $M$:
\begin{eqnarray}
K &= F(J) + F(-J) + 2F(0), \nonumber\\
L &= F(J) - F(-J),         \nonumber\\
M &= F(J) + F(-J) - 2F(0),
\end{eqnarray}
where the function $F(x)$ is defined as
\begin{eqnarray}
F(x) = -\frac{1}{4}\frac{\sum\limits_{n=-s}^{s}n\sinh\left[\beta n(x - B)\right]}{\sum\limits_{n=-s}^{s}\cosh\left[\beta n(x - B)\right]}.
\end{eqnarray}
In view of this notation, the total magnetization $m_s^{+}$ and the staggered magnetization $m_s^{-}$ of the decorating spins normalized per one nodal site of the decorated lattice can be expressed as
\begin{eqnarray}
m_s^{+} &=& \frac{1}{2}\left(m_{h} + m_{v}\right) = K + 4c_I M,
\\
m_s^{-} &=& \frac{1}{2}\left(m_{h} - m_{v}\right) = -4m_I L.
\end{eqnarray}
The other important thermodynamic quantities, such as the Gibbs free energy ${\cal G}$, the entropy $S$ and the specific heat $C$ can easily be obtained from the relations:
\begin{eqnarray}
\label{eq:G}
{\cal G} &=& -k_{\rm B}T\ln {\cal Z}_{I} - 2Nk_{\rm B}T\ln A, \\
\label{eq:S,C}
S &=& -\left(\frac{\partial {\cal G}}{\partial T}\right)_B,
\quad
C = -T\left(\frac{\partial^2 {\cal G}}{\partial T^2}\right)_B.
\end{eqnarray}
Finally, let us make a few comments on a critical behavior of the model.  It is clear from the mapping relation~(\ref{eq:Z}) that the generalized Fisher's super-exchange model on the decorated square lattice may exhibit a critical point only if the corresponding spin-$1/2$ Ising model on the undecorated square lattice is at a critical point, as well. As a consequence, the critical temperature of the mixed spin-$(1/2,s)$ Fisher's super-exchange model can be straightforwardly obtained by comparing the effective nearest-neighbor coupling of the corresponding spin-$1/2$ Ising model on the simple square lattice with its critical value~\cite{Ons44}:
\begin{eqnarray}
\label{eq:Tc}
\beta_c J_{eff} = 2\ln(1+\sqrt{2}),
\end{eqnarray}
where $\beta_c = 1/(k_{\rm B}T_c)$ and $T_c$ denotes the critical temperature of the studied spin model.

\section{Ground-state and finite-temperature properties}
\label{sec:3}

In this section, we present the most interesting numerical results for the ground state, the finite-temperature phase diagram as well as thermal dependencies of the magnetization, entropy and specific heat of the mixed spin-$(1/2, s)$  Fisher's super-exchange model on the decorated square lattice.

First, let us start with a brief description of the ground-state behavior~\cite{note2}. At zero temperature, the investigated spin model passes from the long-range ordered ground state to the paramagnetic one when the magnetic field applied on decorating spins exceeds the critical value $B_c/J = 1$. The former ground state is characterized by a~perfect antiferromagnetic arrangement of the decorating spins placed on horizontal and vertical bonds of the lattice ($m_h = s$, $m_v = -s$) and by the saturated spontaneous magnetization $m_\sigma = 1/2$ attributed to nodal spins. In the latter ground state, all decorating spins are fully polarized towards the magnetic-field direction, while nodal spins are frustrated due to the mutual competition between ferromagnetic and antiferromagnetic exchange interactions ($m_h = m_v = s$, $m_\sigma = 0$).

\begin{figure}[b!]
\vspace{-5mm}
\centering
  \includegraphics[width=0.55\textwidth]{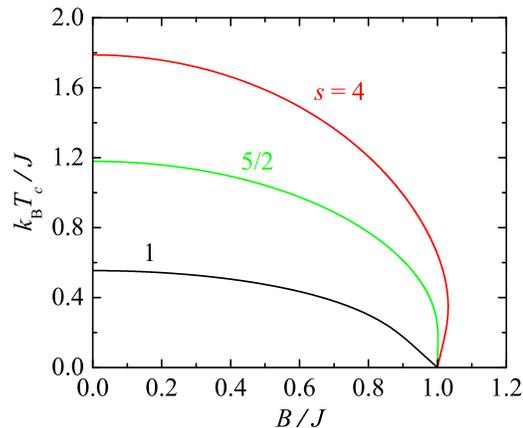}
\vspace{-2mm}
\caption{The finite-temperature phase diagram of the spin-$(1/2, s)$ Fisher's super-exchange model for three different values of decorating spins.}
\label{fig:2}
\end{figure}
As expected, at finite temperatures, the long-range antiferromagnetic order of the decorating spins completely vanishes at the critical temperature given by Eq.~(\ref{eq:Tc}). For better illustration, the critical temperature versus magnetic field is displayed in Fig.~\ref{fig:2} for three different values of the decorating spins. Note that the plotted curves are unique solutions of the critical condition~(\ref{eq:Tc}) and therefore, they present the lines of continuous (second-order) phase transitions between the long-range ordered and paramagnetic phases. As one can see from Fig.~\ref{fig:2}, the critical temperature of the model generally decreases with the increasing magnetic field until it entirely tends to zero at the critical field $B_c/J = 1$ of the first-order phase transition between long-range ordered and paramagnetic ground states. For the decorating spins $s<5/2$, critical lines approach the first-order phase transition with negative slopes, while for reverse case $s\geq5/2$, they approach the critical field $B_c/J = 1$  with positive slopes. These observations clearly suggest that reentrant phase transitions appear in the  magnetic-field region $B/J \gtrsim 1$ just for sufficiently high decorating spins $s\geq5/2$. As can be expected, the observed reentrant phenomenon becomes more pronounced, the higher the spin value $s$ is.

\begin{figure}[t!]
\centering
  \includegraphics[width=0.55\textwidth]{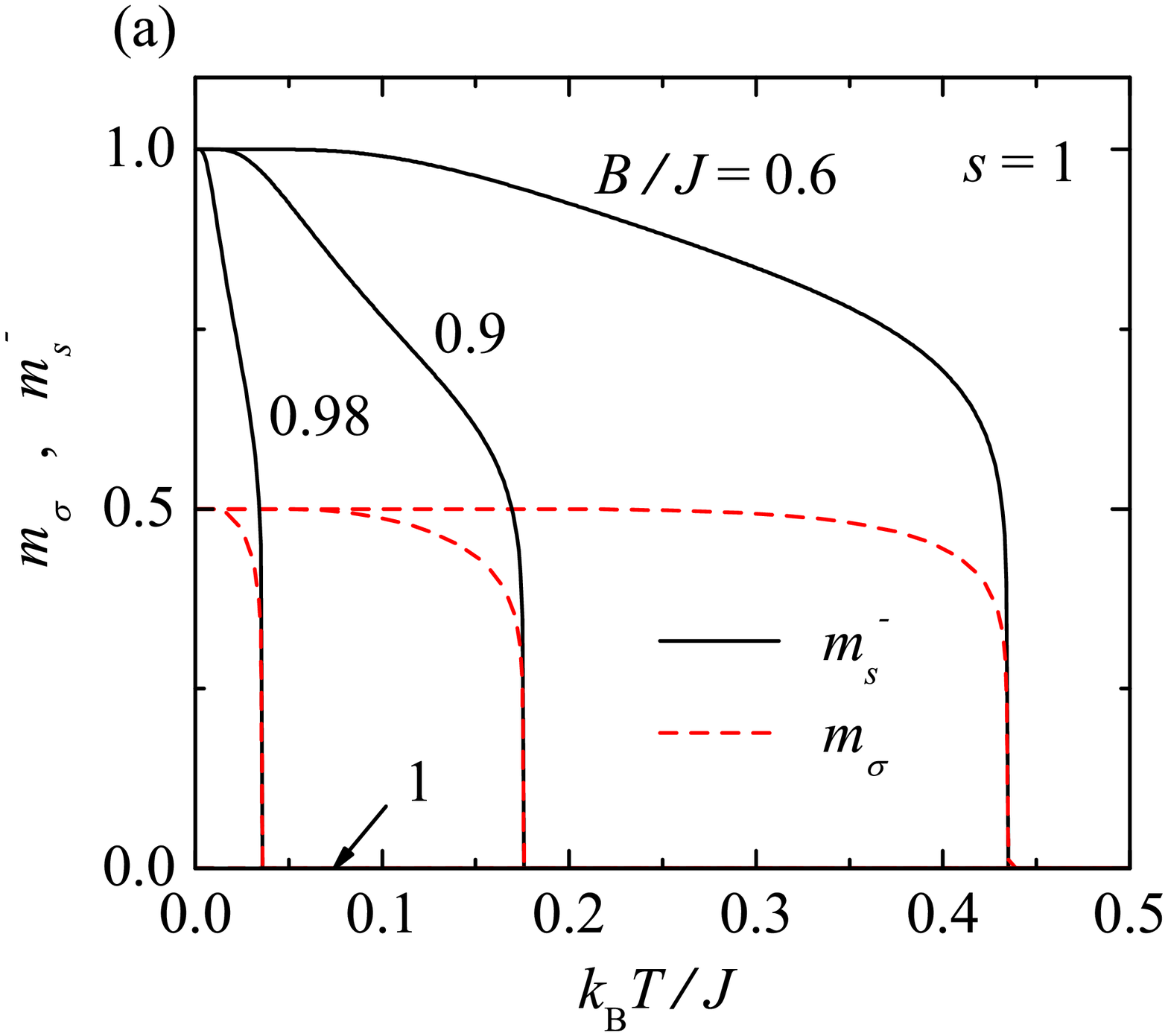}
	\hspace{-2cm}
  \includegraphics[width=0.55\textwidth]{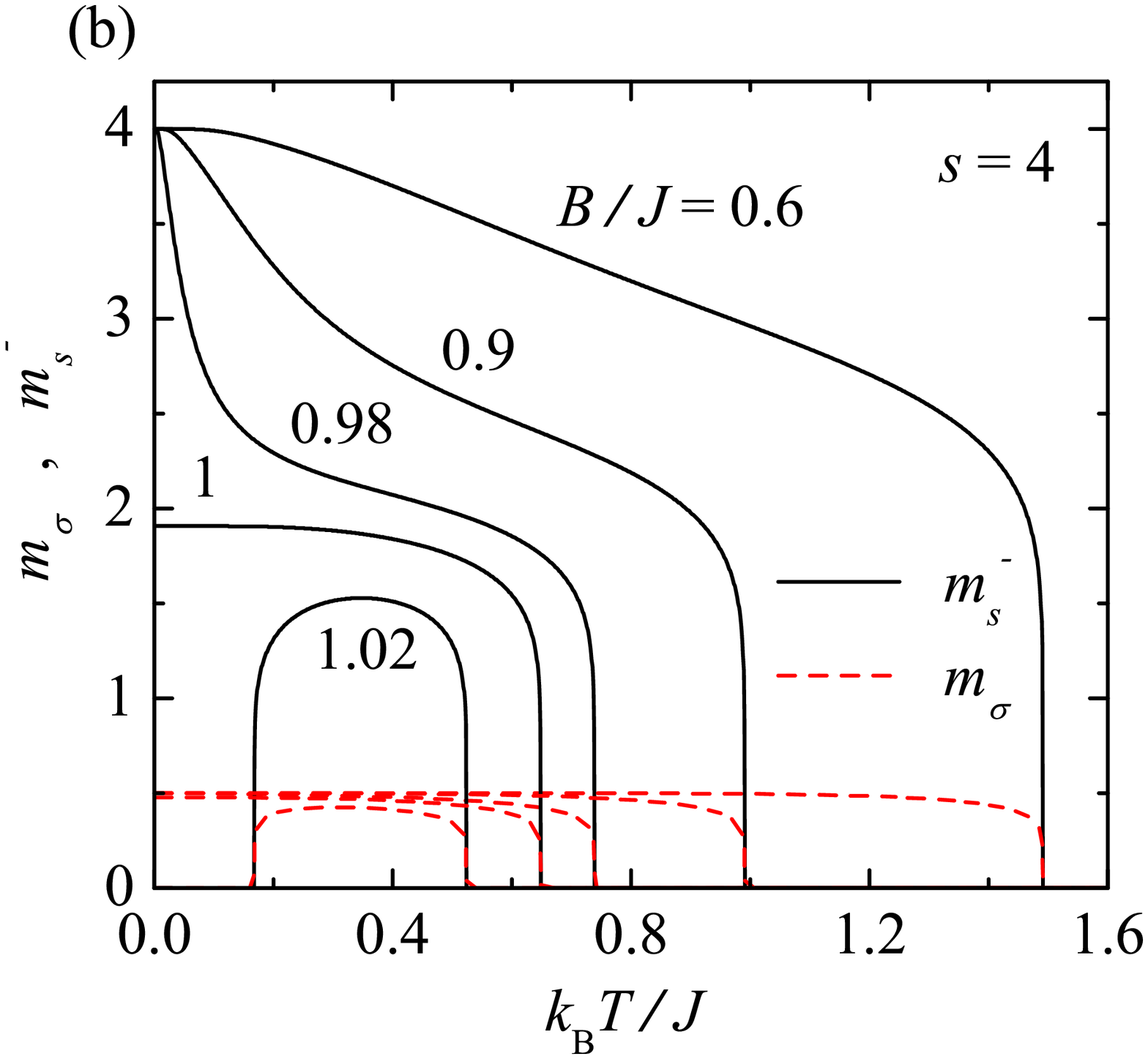}
\vspace{-2mm}
\caption{Temperature variations of the spontaneous magnetization $m_\sigma$ of the nodal spins (broken lines) and the staggered magnetization $m_s^{-}$ of the decorating spins (solid lines) for two particular spin cases (a)~$s = 1$ and (b)~$s = 4$ by assuming a few fixed values of the external magnetic field $B$.}
\label{fig:3}
\end{figure}

To confirm above findings, the temperature dependencies of the spontaneous magnetization $m_\sigma$ of the nodal spins (broken lines) and the staggered magnetization $m_s^{-}$ of the decorating spins are plotted in Fig.~\ref{fig:3} for two particular spin values $s = 1$ and $s = 4$ by assuming different values of the external magnetic field applied on these spins. For easy reference, we will further use the extended N\'eel's classification of $m(T)$ curves~\cite{Nee48, Chi97, Str06}. As one can see from Fig.~\ref{fig:3}, both magnetization start from their saturated values $m_\sigma = 1/2$ and $m_s^{-} = s$ if the applied magnetic field is lower than the critical value $B_c /J = 1$.  Moreover, the spontaneous magnetization of the nodal spins exhibits solely familiar Q-type dependencies characterized by a steep decrease of the magnetization just in the vicinity of critical temperature (see the $m_\sigma(T)$ curves plotted for $B/J=0.6, 0.9, 0.98$ in Figs.~\ref{fig:3}(a) and~(b)). By contrast, temperature dependencies of the staggered magnetization of the decorating spins may change from conventional R-type curves to more interesting S-type curves if values of the decorating spins are high enough and the external magnetic field takes the values $B\lesssim B_c$ (see the $m_s^{-}(T)$ curves corresponding to $B/J=0.6, 0.9$ and $0.98$ in Fig.~\ref{fig:3}(b)). The R-type dependencies exhibit a relatively rapid decline of the magnetization within the range of intermediate temperatures before a sharp drop to zero magnetization at the critical point. The S-type dependencies show two sharp magnetization decreases; the first one, that can be observed at low temperatures, almost completely diminishes in the range of intermediate temperatures and the second one is located nearby the critical temperature. The origin of all three types of the magnetization curves closely relates to the fact that the longitudinal magnetic field $B$ does not directly act on nodal spins, but only on decorating spins localized at horizontal and vertical bonds of the lattice. Hence, the spontaneous magnetization $m_\sigma$ varies very smoothly with temperature, while the staggered magnetization $m_s^{-}$ declines more rapidly as the temperature increases before reaching the critical point. As expected, the observed temperature decrease of $m_s^{-}$ is more rapid, the closer to the critical field $B_c /J = 1$ we are. For the particular case $B_c /J = 1$, the magnetization $m_\sigma$ and $m_s^{-}$ acquire zero-temperature asymptotic values unambiguously given by the general conditions
\begin{eqnarray}
\label{eq:m_sigma_s}
m_\sigma(T=0) &=&
\cases{
0
&for $s<5/2$\,,\\
\frac{1}{2}\left[1  - \frac{(2s+1)^2}{s^4}\right]^{1/8}
&for\, $s\geq5/2$\,,}
\\
m_s^{-}(T=0) &=& s\cdot\, m_\sigma(T=0),
\end{eqnarray}
as illustrated in Fig.~\ref{fig:3}. The above analytical expressions for $m_\sigma$, $m_s^{-}$ indicate that the investigated spin-$(1/2, s)$ Fisher's super-exchange model exhibits an interesting macroscopic degeneracy at $B_c /J = 1$, which originates from the mutual interplay between the magnetic field applied on decorating spins and the ferromagnetic as well as antiferromagnetic exchange interactions between the nearest-neighboring spins in the horizontal and vertical directions, respectively.
Finally, by considering sufficiently high values of the decorating spins $s\geq5/2$, one can observe the reentrant behavior in thermal variations of both the magnetization $m_\sigma$, $m_s^{-}$ with two consecutive critical points for the magnetic fields $B /J \gtrsim 1$ (see the curves plotted for $B /J = 1.02$ in Fig.~\ref{fig:3}(b)), which unambiguously confirms the former analysis of the critical behavior of the studied model.

\begin{figure}[t!]
\centering
  \includegraphics[width=0.55\textwidth]{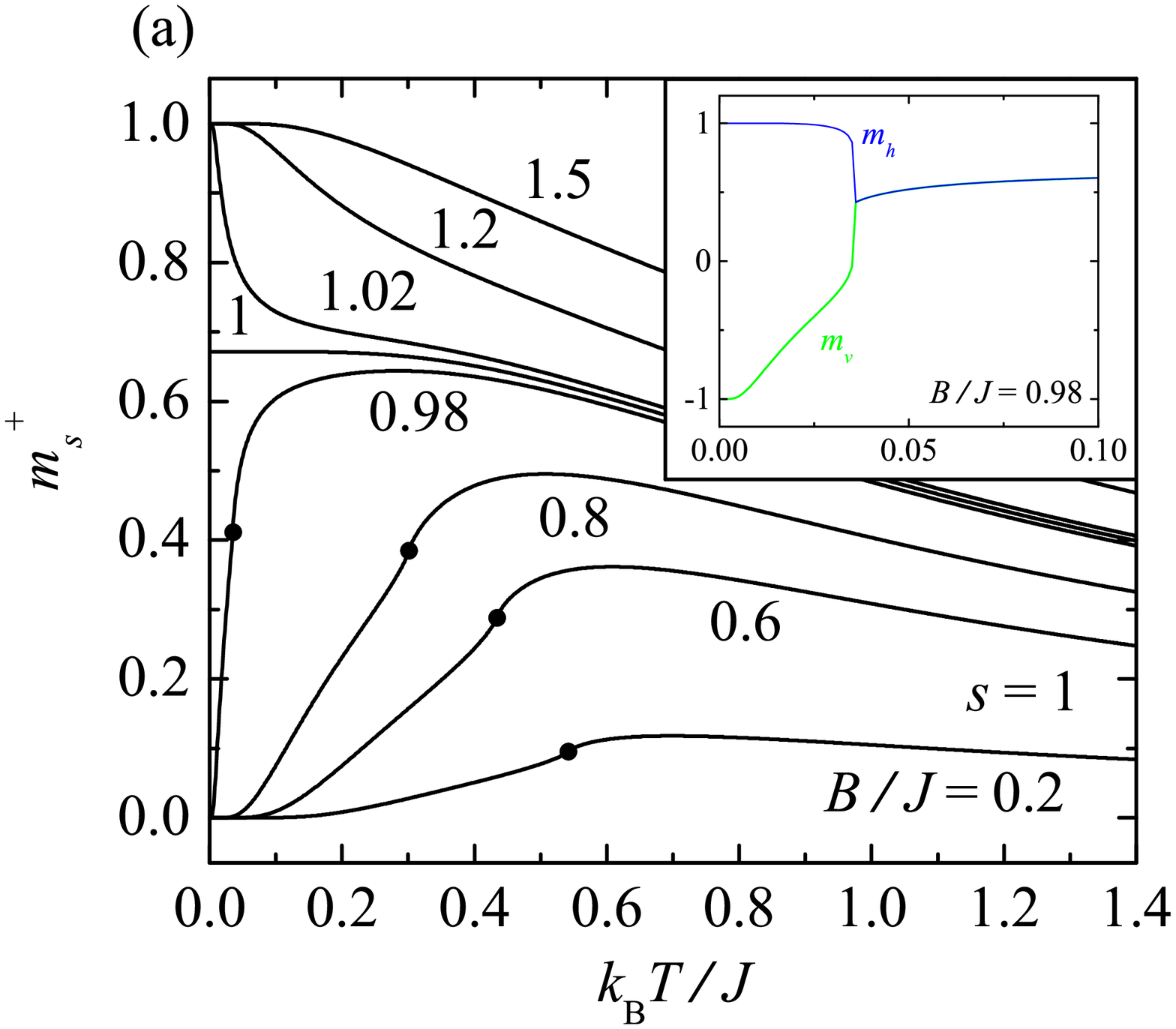}
  \hspace{-2.0cm}
  \includegraphics[width=0.55\textwidth]{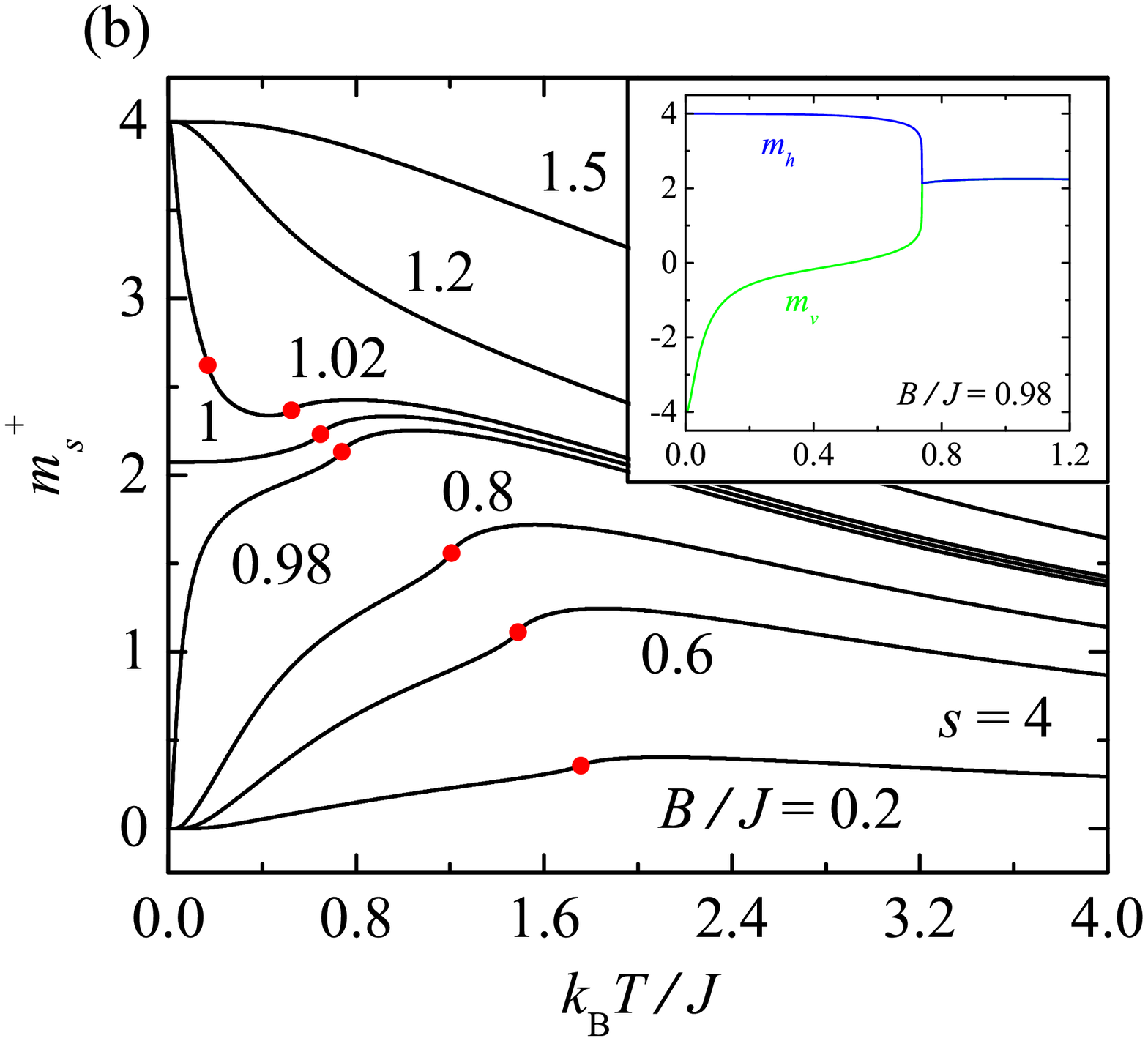}
\vspace{-2mm}
\caption{Temperature variations of the total magnetization $m_s^{+}$ for the same values of decorating spins as in Fig.~\ref{fig:3} by assuming several fixed values of the magnetic field~$B$. Filled circles denote weak energy-type singularities of $m_s^{+}$ at critical temperatures. Insets: Low-temperature variations of the magnetization $m_h$, $m_v$ for the magnetic field $B/J=0.98$.}
\label{fig:4}
\end{figure}

In order to complete the analysis of the magnetization, let us turn our attention to  temperature dependencies of the magnetization $m_s^{+}$, $m_h$ and $m_v$ that are illustrated in Fig.~\ref{fig:4}. The following general conclusions can be deduced from the plotted $m_s^{+}(T)$ curves. Depending on the intensity of applied magnetic field, the total magnetization $m_s^{+}$ can asymptotically reach three different values as the temperature tends to zero, namely,
\begin{eqnarray}
\label{eq:m_s+}
m_s^{+}(T=0) =
\cases{
0
 &for $B < B_c$\,,
\\
\frac{s-2}{4} + \frac{2s+1}{\pi(s+1)}\,{\cal K}\left(\frac{s\sqrt{2s+1}}{s+1}\right)
&for $B = B_c$\,,
\\
s
& for $B > B_c$\,,
}
\end{eqnarray}
where ${\cal K}(x) = \int_0^{\pi/2}\left(1-x^2\sin^2\phi\right)^{-1/2}{\rm d}\phi$ is a complete elliptic integral of the first kind. In the low-temperature region, $m_s^{+}$ shows a noticeable increase (decrease) with increasing temperature if the magnetic field takes lower (higher) values than $B_c/J=1$. As expected, these temperature-induced changes of $m_s^{+}$ are more pronounced, the closer the magnetic field is to the critical value $B_c/J=1$. The rapid variations of  $m_s^{+}$ observed in the temperature regime $T<T_c(B\neq0)$ are evidently associated with predominant temperature-induced excitations of the decorating spins located at vertical bonds of the lattice, which are clearly reflected also in an unusual steep low-temperature variation of the corresponding magnetization $m_v$ (see the insets in Figs.~\ref{fig:4}(a) and~\ref{fig:4}(b)).
Moreover, one or two weak energy-type singularities can also be found in $m_s^{+}(T)$ curves at critical temperatures relevant to continuous phase transitions between the long-range ordered state and the paramagnetic state in dependence on the magnitude of the decorating spins and the intensity of the magnetic field applied on these spins. One can also see from Fig.~\ref{fig:4}, where these singularities are denoted by full circles, that the magnetization $m_s^{+}$ exhibits an interesting broad local maximum above the critical temperature, at which the studied spin system passes from the long-range ordered state to the paramagnetic state as the temperature increases. As demonstrated by M. E. Fisher~\cite{Fis60a}, the intriguing temperature-induced increase of $m_s^{+}$ indicates the presence of the residual short-range ordering in the the temperature region $T\gtrsim T_c(B\neq0)$. On the other hand, if decorating spins are high enough to form the reentrant critical behavior in the field region $B_c/J\gtrsim1$, then, a sharp drop in temperature dependencies of $m_s^{+}$ resulting to a local minimum can be detected in the relatively narrow temperature range between two successive singularities (see the curve plotted in Fig.~\ref{fig:4}(b) for the magnetic field $B/J=1.02$).
\begin{figure}[t!]
\centering
  \includegraphics[width=0.55\columnwidth]{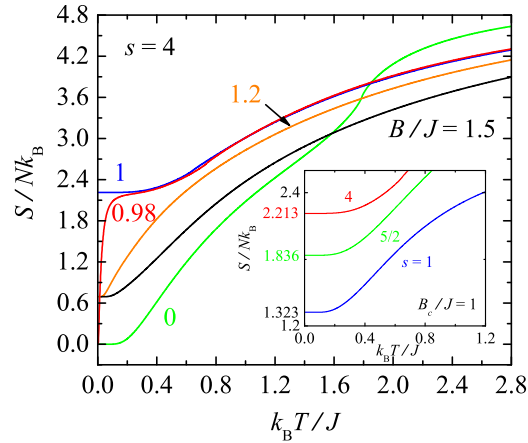}
\vspace{-2mm}
\caption{Temperature variations of the entropy for the spin case $s = 4$ and several fixed values of the magnetic field $B$. Inset: Low-temperature variations of the entropy for three selected values of decorating spins by assuming the critical field $B_c/J=1$.}
\label{fig:5}
\end{figure}

Now, let us look in detail at temperature variations of basic thermodynamic quantities such as the entropy and specific heat. Figure~\ref{fig:5} shows temperature dependencies of the entropy normalized per one nodal site of the decorated square lattice calculated for one representative spin value $s=4$ and a few different values of the magnetic field $B$ (main figure) as well as for three different values of decorating spins by assuming the critical field $B_c/J =1$ (inset). As one can see, the displayed entropy curves tend asymptotically either to zero or to the finite value $S(T=0)=Nk_{\rm B}\ln 2\approx 0.693Nk_{\rm B}$ in the zero-temperature limit $T\to 0$ depending on whether $B<B_c$ or $B>B_c$, respectively. In accordance with the ground-state analysis, the origin of the zero-temperature residual entropy $S(T=0)=Nk_{\rm B}\ln 2$ lies solely in the spin frustration of the nodal spins $\sigma$ observed within the paramagnetic ground state. Hence, it remains the same for any magnetic fields $B>B_c$ regardless of the magnitude~$s$. In contrast to this, as far as the critical field $B_c/J=1$ is considered, the entropy of the system reaches the highly non-trivial asymptotic value $S_c(T=0)>Nk_{\rm B}\ln (2s+1)\geq Nk_{\rm B}\ln 2$ at zero temperature which increases with the increasing value of decorating spins (see the inset in Fig.~\ref{fig:5}). The origin of this value can not be explained by any simple argument. It is possible just say that it is determined by the cooperative action of the whole lattice.

\begin{figure}[t!]
\centering
  \includegraphics[width=0.55\columnwidth]{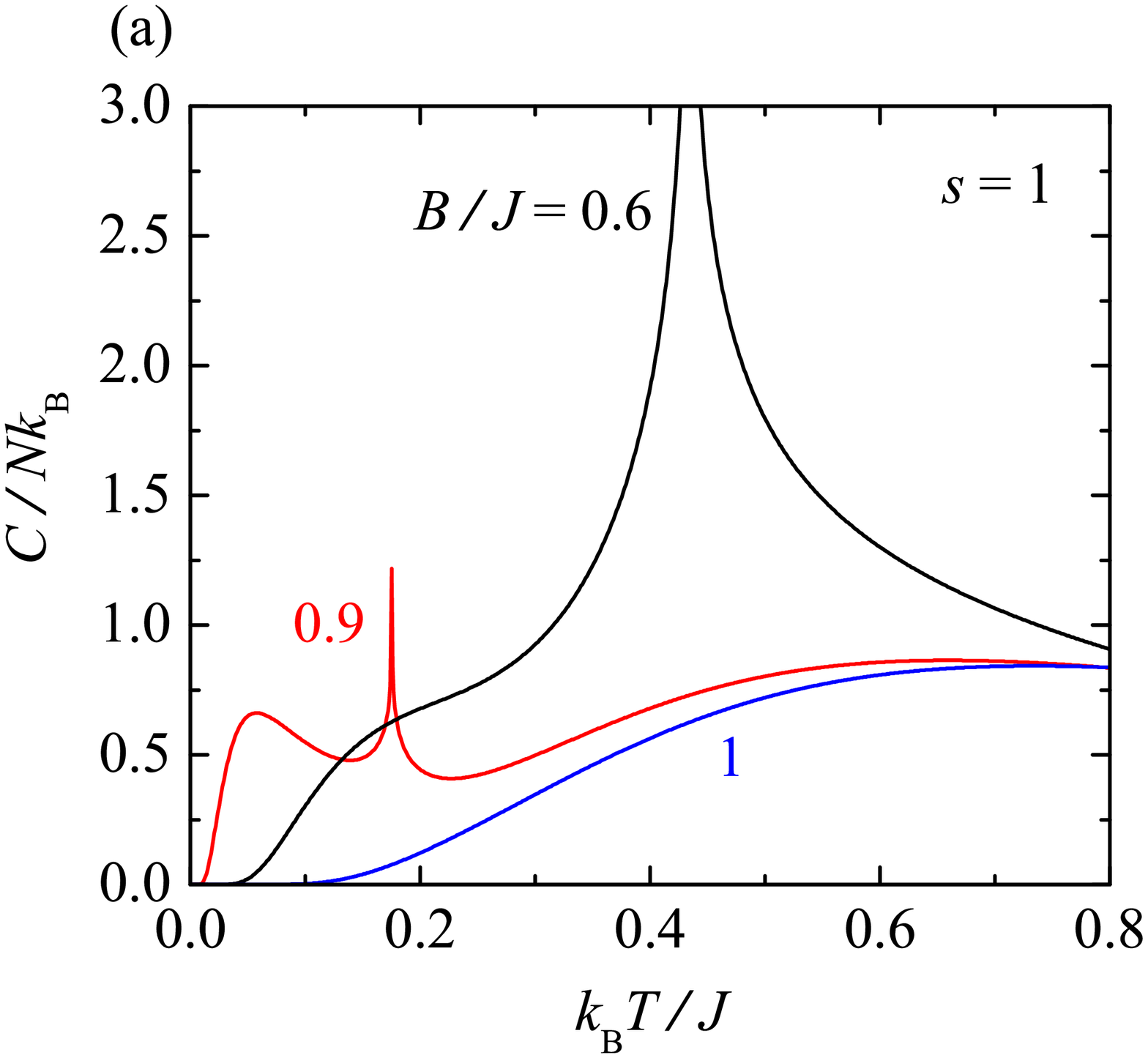}
  \hspace{-2cm}
  \includegraphics[width=0.55\columnwidth]{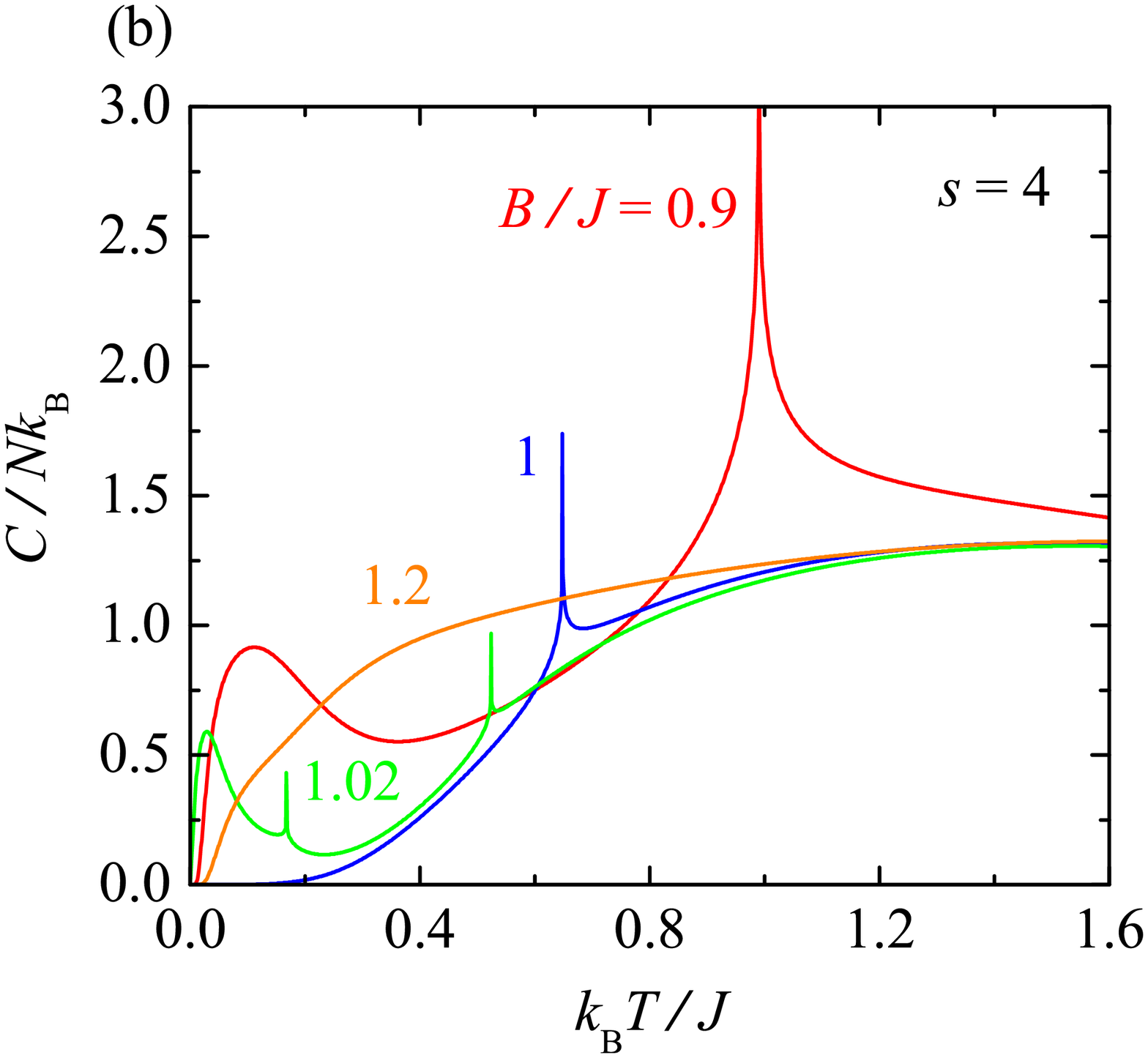}
\vspace{-2mm}
\caption{Temperature variations of the specific heat for the same decorating spins as in Figs.~\ref{fig:3} and \ref{fig:4} by considering a few fixed values of the magnetic field~$B$.}
\label{fig:6}
\end{figure}

Finally, we conclude the analysis of thermodynamics with a description of typical temperature dependencies of the specific heat that are displayed in Fig.~\ref{fig:6}. To enable a direct comparison, we have chosen the values of the external magnetic field and decorating spins so as to match the finite-temperature phase diagram shown in Fig.~\ref{fig:2} and also temperature dependencies of the magnatization plotted in Figs.~\ref{fig:3} and~\ref{fig:4}. In this manner, depicted specific heat curves reflect a comprehensive picture of the finite-temperature behavior of the investigated spin system. In fact, besides the one or two logarithmic divergences at appropriate critical temperatures, marked local maxima can be detected in low-temperature tails of the specific heat curves if the external magnetic field takes the values from a close vicinity of the first-order phase transition between the long-range ordered ground state and the paramagnetic ground state. A direct comparison of the specific heat curves depicted in Fig.~\ref{fig:6} for the magnetic fields $B/J=0.9$ and $1.02$ with the corresponding magnetization $m_\sigma$, $m_s^{-}$, $m_s^{+}$ shown in Figs.~\ref{fig:3} and~\ref{fig:4} confirms that the origin of observed low-temperature maxima lies in strong thermal excitations to a spin configuration rather close in energy to the ground state.

\section{Magnetocaloric properties}
\label{sec:4}

Since the investigated spin-$(1/2, s)$ Fisher's super-exchange model on a decorated square lattice~(\ref{eq:H}) is exactly solvable within the generalized decoration-iteration mapping
transformation~\cite{Fis59, Syo72, Str10}, it provides an excellent paradigmatic example of an exactly soluble two-dimensional spin system, which allows an examination of the MCE in a~vicinity of the continuous phase transition at finite
magnetic fields. Actually, the magnetocaloric quantities,
such as the isothermal entropy change $\Delta S_T$ and the
adiabatic temperature change $\Delta T_{ad}$ upon the magnetic-field
variation $\Delta B\!: 0 \to B$ can be rigorously calculated
by using the following formulas:
\begin{eqnarray}
\label{eq:dS}
\Delta S_T(T, \Delta B) &=  S(T, B\neq 0) - S(T, B= 0), \\
\label{eq:dT}
\Delta T_{ad}(S, \Delta B) &=  T(S, B\neq 0) - T(S, B= 0).
\end{eqnarray}
Recall that the former relation~(\ref{eq:dS}) is valid if the temperature $T$ of the model is constant, while the latter relation~(\ref{eq:dT}) satisfies the adiabatic condition \mbox{$S = {\rm const.}$}.

\begin{figure}[t!]
\centering
  \includegraphics[width=0.55\textwidth]{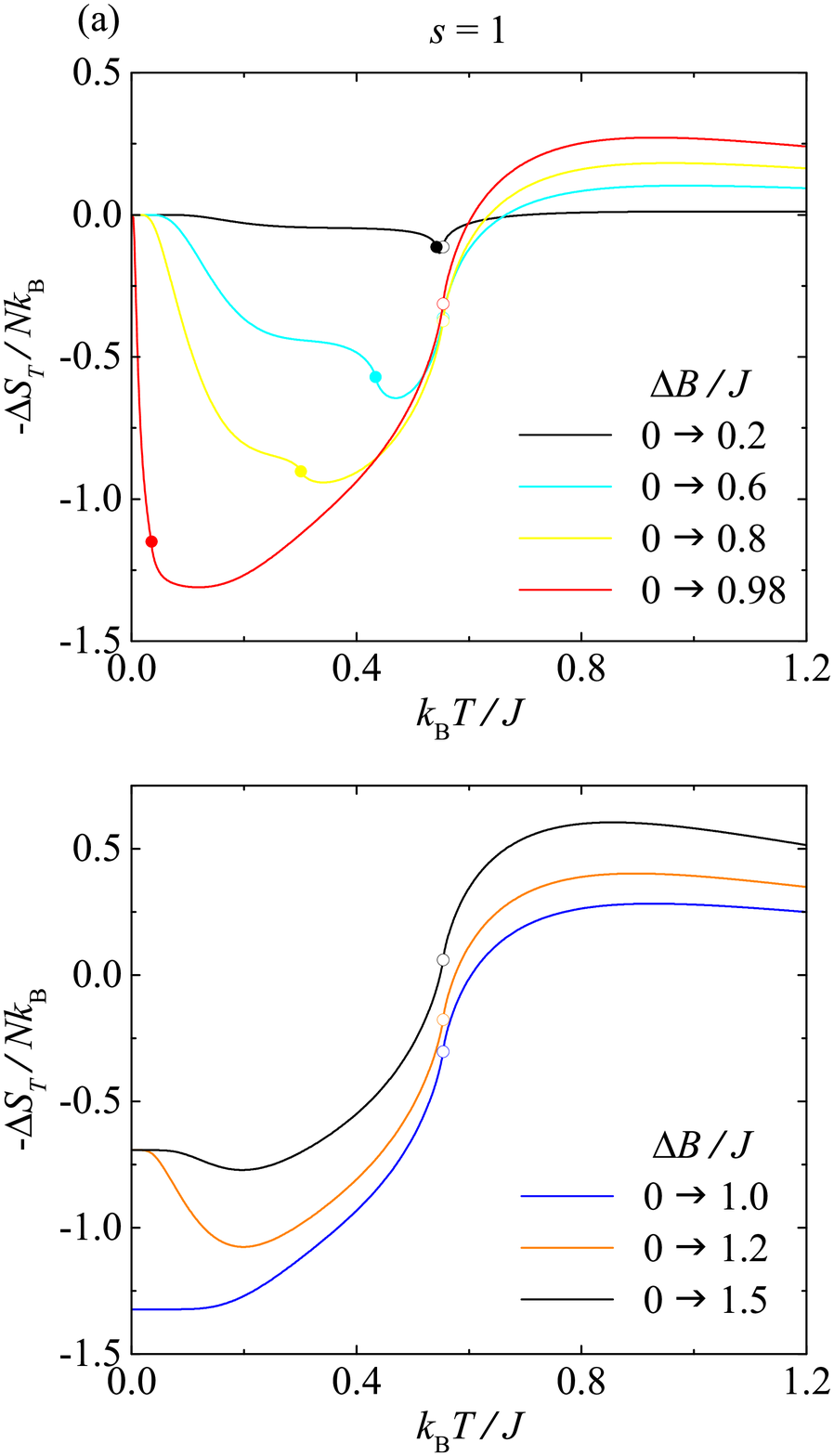}
   \hspace{-1.85cm}
  \includegraphics[width=0.55\textwidth]{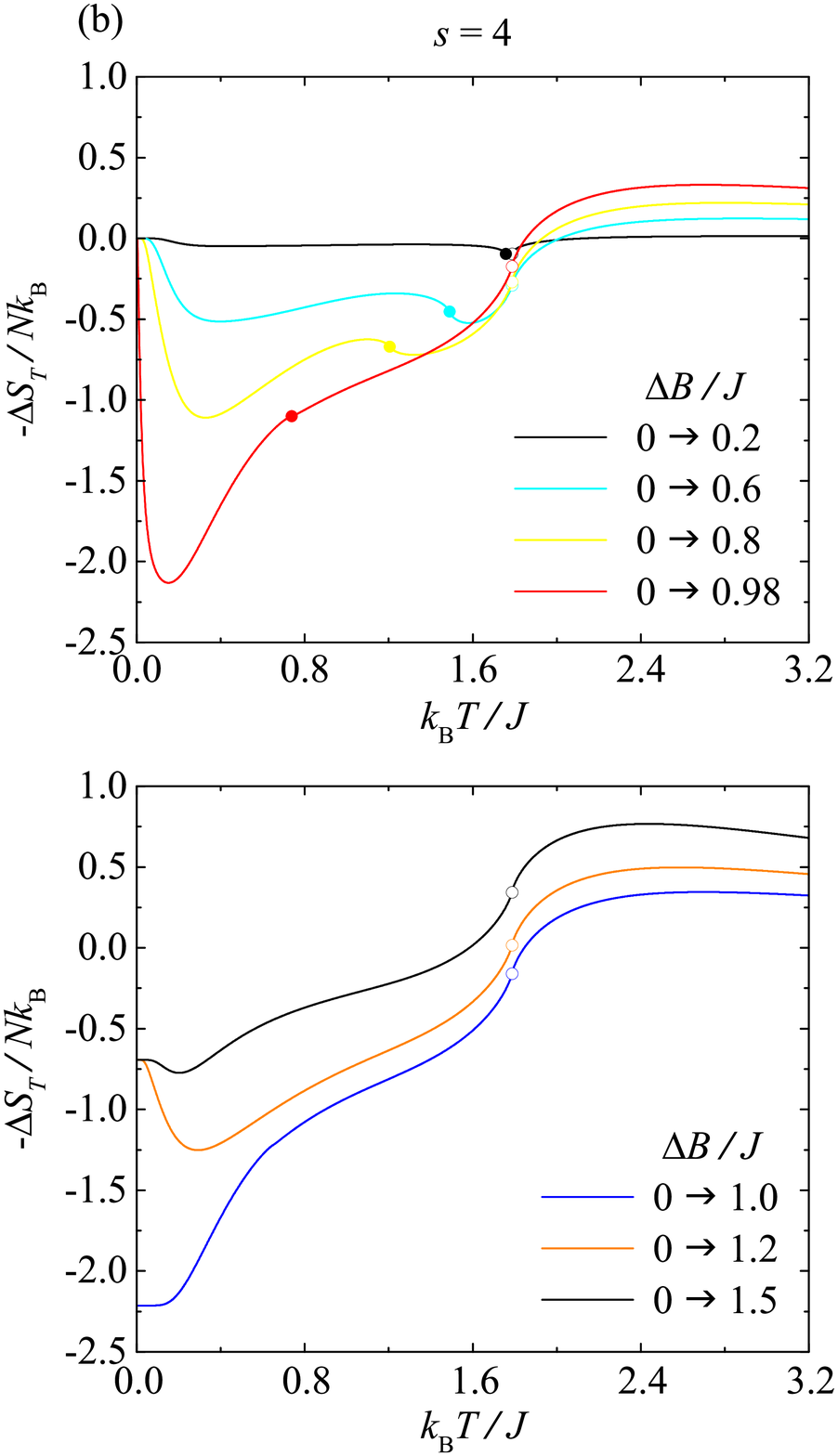}	
\vspace{-10mm}
\caption{Temperature variations of the isothermal entropy change normalized per one nodal lattice site ($-\Delta S_T/Nk_{\rm B}$) for the same decorating spins as in Figs.~\ref{fig:3}--\ref{fig:6} by considering few magnetic-field changes $\Delta B\!: 0 \to B$. Empty and full circles mark weak singularities of the entropy located at critical points of continuous phase transitions at zero and respective non-zero magnetic fields, respectively.}
\label{fig:7}
\end{figure}

Figure~\ref{fig:7} shows temperature dependencies of the isothermal entropy change normalized per site of the original square lattice ($-\Delta S_{T}/Nk_{\rm B}$) for two particular values of decorating spins $s=1$ and $s=4$ by considering various magnetic-field changes $\Delta B\!: 0\to B$. As one can see, the isothermal entropy change may be either positive or negative depending on the temperature, which clearly points to both conventional ($-\Delta S_{T}>0$) and inverse ($-\Delta S_{T}<0$) MCE for any magnetic-field change $\Delta B$. Namely, in the high-temperature region $T\gg T_c(B=0)$, where only short-range ordering occurs, $-\Delta S_{T}$ is always positive and slowly increases to the broad maximum with decreasing temperature due to suppression of the spin disorder by the applied magnetic field. At certain temperature, $-\Delta S_{T}$ starts to rapidly decrease and changes sign from positive to negative as the temperature further decreases. Rounded negative minima in low-temperature parts of $-\Delta S_{T}(T)$ curves detected at the temperatures $T < T_c(B=0)$ for magnetic-field changes $\Delta B/J\in\left(0, 1\right)$ clearly indicate the presence of an enhanced inverse MCE in this temperature region. To be more specific, for the decorating spins $s\leq 1$, the local minimum occurs merely in the temperature interval $T_c(B\neq0) \lesssim T < T_c(B=0)$, i.e. slightly above the second-order phase transition (see the upper panel in Fig.~\ref{fig:7}(a)). Evidently, this minimum gradually enlarges and shifts to lower temperatures upon the increase of the field change $\Delta B$. In accordance with these observations, the origin of the detected enhanced inverse MCE can be attributed to strong thermal fluctuations of spins leading to an unusual increase of the total magnetization $m_s^{+}$ in this region (compare $-\Delta S_{T}(T)$ curves plotted in upper panel in Fig.~\ref{fig:7}(a) with the corresponding temperature variations of the magnetization $m_s^{+}$ shown in Fig.~\ref{fig:4}(a)).
More complex scenario occurs if decorating spins take the higher values than $s=1$. Two local minima can be observed in low-temperature parts of $-\Delta S_{T}(T)$ curves for $\Delta B/J\in\left(0, 1\right)$ provided sufficiently high decorating spins. Indeed, one minimum can be detected in the temperature range $T_c(B\neq0) \lesssim T < T_c(B=0)$, while the other one creates below the second-order phase transitions  at the temperatures $T < T_c(B\neq0)$ with the increasing intensity of the applied magnetic field, as shown in the upper panel in Fig.~\ref{fig:7}(b) for the representative spin case $s=4$. Obviously, if the magnetic-field change approaches $\Delta B/J\!:0\to 1$, these two local minima gradually merge into one pronounced minimum located far below the critical temperature $T_c(B\neq0)$ (see the curve corresponding to $\Delta B/J\!:0\to 0.98$ in the upper panel of Fig.~\ref{fig:7}(b)). It is justified to suppose that the enhanced inverse MCE detected below $T_c(B\neq0)$ comes from strong thermal excitations of the decorating spins placed on vertical bonds of the lattice, which are reflected in a sharp temperature-induced increase of the corresponding magnetization $m_{v}$ and, subsequently, also the magnetization $m_{s}^{+}$ (compare the $-\Delta S_{T}(T)$ curve plotted in the upper panel of Fig.~\ref{fig:7}(b) for $\Delta B/J\!:0\to 0.98$ with corresponding temperature dependencies of the magnetization $m_v$ and $m_s^{+}$ displayed in Fig.~\ref{fig:4}(b)). As expected, this inverse MCE enlarges with the increasing spin value $s$ due to the increase of predominant thermal excitations of decorating spins placed on vertical bonds of the lattice (it is not shown). Furthermore, it is quite obvious from Fig.~\ref{fig:7} that $-\Delta S_{T}(T,\Delta B\!: 0\to 1)<-\Delta S_{T}(T,\Delta B\!:0\to B\neq 1)$ is satisfied if the temperature approaches the zero value. Thus, one may conclude that the most pronounced inverse MCE can always be found for the magnetic-field change $\Delta B/J\!: 0\to 1$, which exactly coincides with the critical field $B_c/J = 1$ of the first-order phase transition between the magnetically ordered and paramagnetic ground states. Finally, for $B/J > 1$, the inverse MCE (minimum in $-\Delta S_{T}(T)$ curves) is gradually reduced with the increasing $\Delta B$ due to weakening of thermal excitations from the paramagnetic ground state towards the long-range ordered excited state (see lower panels in Fig.~\ref{fig:7}).

\begin{figure}[t!]
\centering
  \includegraphics[width=0.55\textwidth]{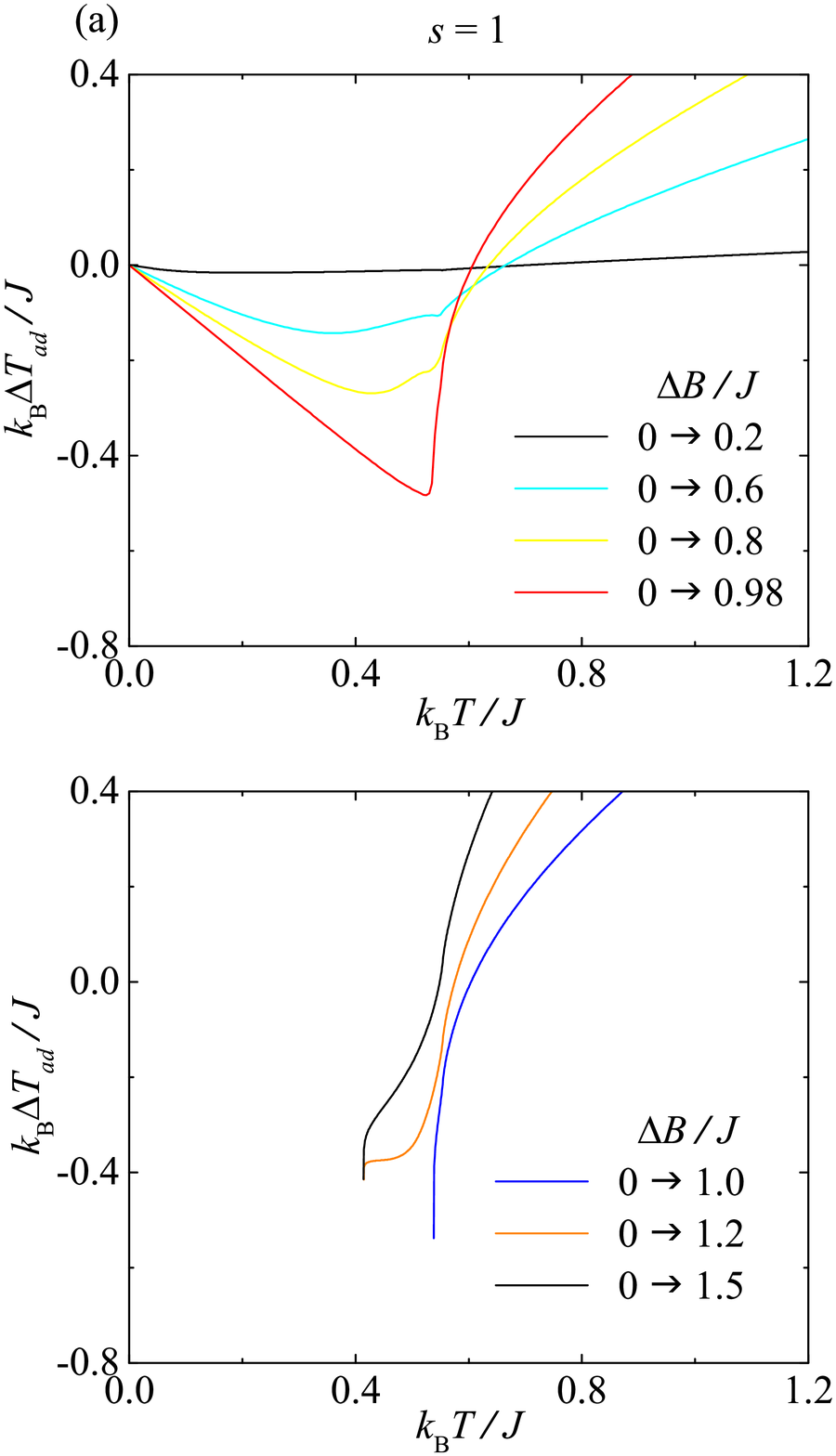}
  \hspace{-1.85cm}
  \includegraphics[width=0.55\textwidth]{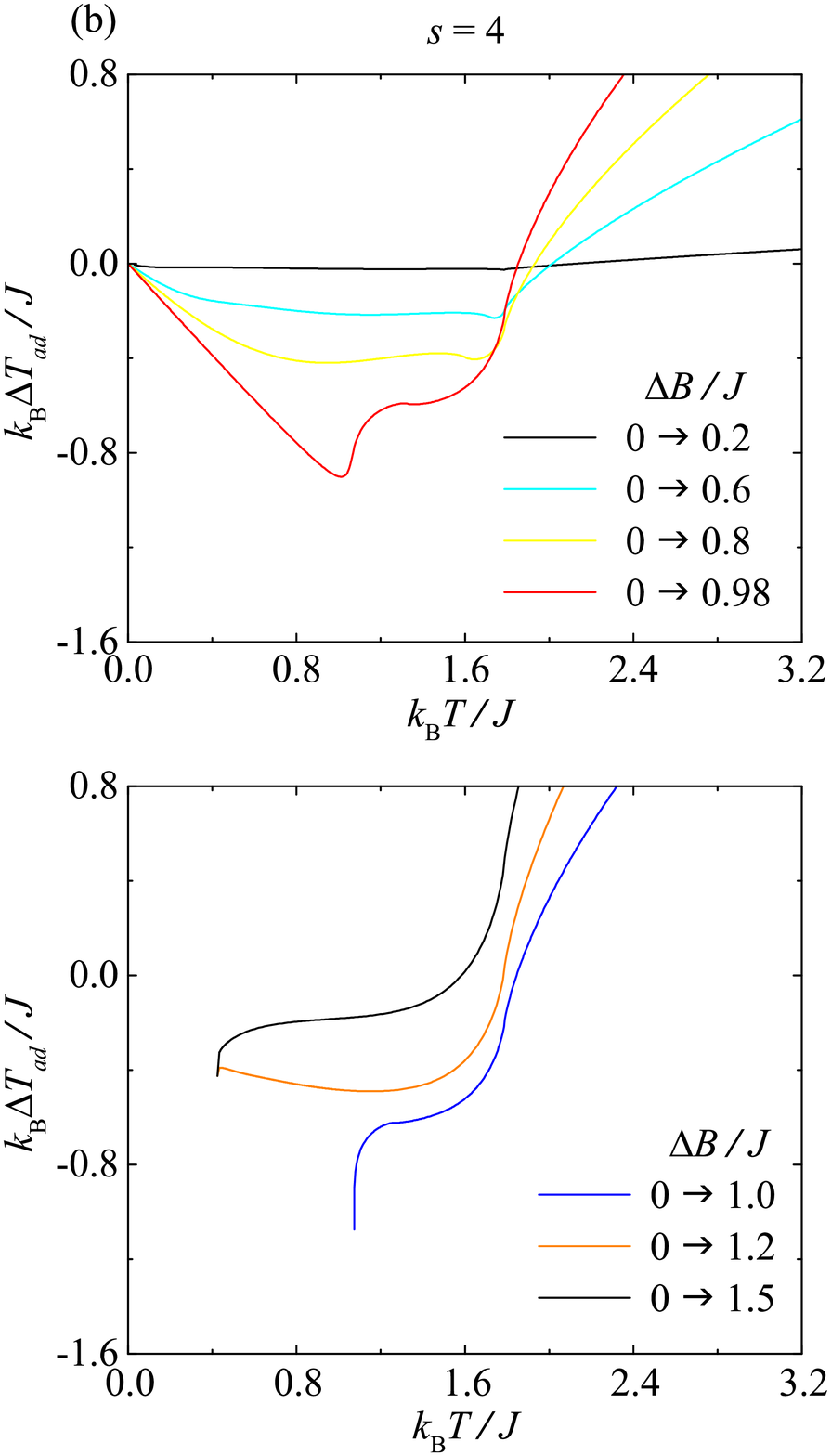}	
\vspace{-10mm}
\caption{Temperature variations of the adiabatic temperature change ($k_{\rm B}\Delta T_{ad}/J$) for the same decorating spins and magnetic-field changes as in Fig.~\ref{fig:7}.}
\label{fig:8}
\end{figure}

To discuss the MCE, one may alternatively examine the adiabatic temperature change $\Delta T_{ad}$ of the system at various magnetic-field changes $\Delta B\!: 0 \to B$. Typical temperature variations of this magnetocaloric potential for the considered model are displayed in Fig.~\ref{fig:8}. Note that all curves plotted in Fig.~\ref{fig:8} were calculated using Eq.~(\ref{eq:dT}) by keeping the entropy constant. As one can see, the adiabatic temperature change $\Delta T_{ad}$ clearly allows to distinguish the conventional MCE ($\Delta T_{ad}>0$) from the inverse MCE ($\Delta T_{ad}<0$). In accordance to the previous discussion, the investigated spin system generally heats up fast as possible in a close vicinity of the first-order phase boundary between long-range ordered and paramagnetic ground states achieved upon the adiabatic reduction of the magnetic field regardless of the magnitude of decorating spins. Indeed, the magnitude of the negative peak in $\Delta T_{ad}(T)$ curves gradually increases with the magnetic-field change $\Delta B$ (see upper panels of Fig.~\ref{fig:8}). In addition, $\Delta T_{ad}$ versus temperature plots end at zero value in the asymptotic limit of zero temperature for any $\Delta B/J\in(0, 1)$, which can be attributed to a perfect antiferromagnetic order of decorating spins placed on horizontal and vertical bonds of the square lattice at zero temperature.
By contrast, the adiabatic temperature change rapidly drops to finite negative values at certain temperatures when the applied magnetic field is equal to or higher than the critical value $B_c/J =1$ (see lower panels in Fig.~\ref{fig:8}).  In this particular case, the magnetocaloric potential $\Delta T_{ad}$ cannot be defined below aforementioned temperatures, because there is no temperature end point in the adiabatic process if $B/J \geq 1$. This intriguing behavior is evidently caused by residual entropies found at the coexistence point $B_c/J = 1$ of the first-order phase transition and within the paramagnetic ground state (see Fig.~\ref{fig:5}).

\section{Summary and future outlooks}
\label{sec:5}

The present work deals with the thermodynamics and magnetocaloric properties of the generalized spin-$(1/2, s)$ Fisher's super-exchange antiferromagnet on the decorated square lattice. Exact results for the critical temperature, total and sublattice magnetization, specific heat and entropy have been derived and discussed in detail for a few representative values of decorating spins. It has been shown that the studied mixed-spin model exhibits reentrant phase transitions with two consecutive critical points slightly above the critical field $B_c/J = 1$ corresponding to the first-order phase transition between the long-range ordered and paramagnetic ground states if decorating spins take the sufficiently high values $s\geq 5/2$. The existence of this non-trivial phenomenon has also been confirmed by temperature variations of the spontaneous and staggered magnetization of the nodal and decorating spins, respectively, as well as, by remarkable temperature dependencies of the specific heat exhibiting two logarithmic singularities.

Moreover, the MCE has been particularly examined by means of the isothermal entropy change and the adiabatic temperature change. The investigation of both magnetocaloric potentials has enabled us to rigorously clarify the magnetic refrigeration efficiency of the considered spin model in a vicinity of the first-order phase transition between the long-range ordered ground state and the paramagnetic ground state as well as nearby the critical temperature, which completely destroys the antiferromagnetic long-range order. The obtained results clearly indicated the fast heating of the studied mixed-spin system during the adiabatic demagnetization process (i.e., a presence of the enhanced inverse MCE) in these regions due to strong thermal spin fluctuations leading to the temperature-induced increase of the total magnetization corresponding to decorating spins. The maximal heating efficiency of the system has been observed for the magnetic-field change $\Delta B/J\!: 0\to 1$, which coincides with the first-order phase transition between the long-range ordered and paramagnetic ground states.

Finally, it is worthwhile to mention that the presented generalization of the Fisher's super-exchange antiferromagnet on a decorated lattice in terms of arbitrary decorating spins is just one of many possible. Other simple generalizations allowing rigorous investigation of an enhanced MCE in two-dimensional spin systems are the introduction of the second-neighbor interaction between nodal spins~\cite{Hat68}, the introduction of the chemical potential~\cite{Mas73}, the axial zero-field splitting~\cite{Can06}, the transverse magnetic field as well as the rhombic zero-field splitting on decorating spins. Moreover, one may also consider other planar lattices with the even coordination number, such as the kagom\'e lattice and the triangular lattice. Our future work will continue in this direction.

\section*{Acknowledgments}
This work was financially supported by Ministry of Education, Science,
Research and Sport of the Slovak Republic provided under the grant VEGA~1/0043/16 and by Slovak Research and Development Agency provided under the contract No.~APVV-0097-12.

\section*{References}
\bibliographystyle{elsarticle-num}

\begin{thebibliography}{50}
\bibitem{War81}
Warburg E 1881 {\it Ann. Phys. (Leipzig)} {\bf 13} 141--64.
\bibitem{Gia33}
Giauque W F and MacDougall D P 1933 \PR {\bf 43} 768.
\bibitem{Str07}
Strehlow P, Nuzha H and Bork E 2007 {\it J. Low Temp. Phys.} {\bf 147} 81--93.
\bibitem{Zhi03}
Zhitomirsky M E 2003 \PR B {\bf 67} 104421.
\bibitem{Zhi04}
Zhitomirsky M E and Honecker A 2004 {\it J. Stat. Mech.: Theor. Exp.} P07012.
\bibitem{Der06}
Derzhko O and Richter J 2006 {\it Eur. Phys. J.} B {\bf  52} 23--36.
\bibitem{Sch07}
Schnack J, Schmidt R and Richter J 2007 \PR B {\bf 76} 054413.
\bibitem{Hon09a}
Honecker A and Zhitomirsky M E 2009 {\it J. Phys.: Conf. Ser.} {\bf 145} 012082.
\bibitem{Hon09}
Honecker A and Wessel S 2009 {\it Condens. Matter Phys.} {\bf 12} 399--410.
\bibitem{Can09}
\v{C}anov\'a L, Stre\v{c}ka J and Lu\v{c}ivjansk\'y T 2009 {\it Condens. Matter Phys.} {\bf 12} 353--68.
\bibitem{Tri10}
Trippe C, Honecker A, Kl\"umper A and Ohanyan V 2010 \PR B {\bf 81} 054402.
\bibitem{Lan10}
Lang M, Tsui Y, Wolf B, Jaiswal-Nagar D, Tutsch U, Honecker A, Removi\'{c}-Langer C, Prokofiev A, Assmus W and Donath G 2010 {\it J. Low Temp. Phys.} {\bf 159} 88--91.
\bibitem{Hon11}
Honecker A, Hu S, Peters R and Richter J 2011 \JPCM {\bf 23} 164211.
\bibitem{Top12}
Topilko M, Krokhmalskii T, Derzhko O and Ohanyan V 2012 {\it Eur. Phys. J.} B {\bf 85} 278.
\bibitem{Ver13}
Verkholyak T and Stre\v{c}ka J 2013 \PR B {\bf 88} 134419.
\bibitem{Kas13}
Kassan-Ogly F A, Medvedev M V, Proshkin A I and Zarubin A V 2013 {\it Bulletin of the Russian Academy of Sciences: Physics} {\bf 77} 1245--7.
\bibitem{Gal14}
G\'alisov\'a L 2014 {\it Condens. Matter Phys.} {\bf 17} 13001.
\bibitem{Str14a}
Stre\v{c}ka J, Rojas O, Verkholyak T and Lyra M L 2014 \PR E {\bf 89} (2014) 022143.
\bibitem{Zar15}
Zarubin A V, Kassan-Ogly F A, Medvedev M V and Proshkin A I 2015 {\it Condens. Solid State Phenomena} {\bf 233--234} 212.
\bibitem{Gal15a}
G\'alisov\'a L 2015 {\it Acta Mechanica Slovaca} {\bf 19} 46--53.
\bibitem{Per09}
Pereira M S S, de Moura F A B F and Lyra M L 2009 \PR B {\bf 79} 054427.
\bibitem{Gal15b}
G\'alisov\'a L and Stre\v{c}ka J 2015 {\it Acta Phys. Pol.} A {\bf 127} 216--8.
\bibitem{Gal15c}
G\'alisov\'a L and Stre\v{c}ka J 2015 \PR E {\bf 91} 022134.
\bibitem{Gal15d}
G\'alisov\'a L and Stre\v{c}ka J 2015 \PL A  {\bf 379} 2474--8.
\bibitem{Str14b}
Stre\v{c}ka J and \v{C}is\'arov\'a J 2014 {\it Acta Phys. Pol.} A {\bf 126} 26--7.
\bibitem{Sha14}
Sharples J W, Collison D, McInnes E J L, Schnack J, Palacios E and Evangelisti M 2014 {\it Nature Communications} {\bf 5} 5321.
\bibitem{Str15a}
Stre\v{c}ka J, Kar\v{l}ov\'a K and Madaras T 2015 {\it Physica} B {\bf 466--467} 76--85.
\bibitem{Sza14}
Sza{\l}owski K and Balcerzak T 2014 {\it J. Phys.: Condens. Matter} {\bf 26} 386003.
\bibitem{Hu08}
Hu Y and Du A 2008 {\it J. Phys.: Condens. Matter} {\bf 20} 125225.
\bibitem{Oli08}
de Oliveira N A and von Ranke P J 2008 \PR B {\bf 77} 214439.
\bibitem{Ran10}
von Ranke P J, de Oliveira N A, Alho B P, de Sousa V S R, Plaza E J R, Magnus A and Carvalho~G 2010 \JMMM {\bf 322} 84.
\bibitem{Sza11}
Sza{\l}owski K, Balcerzak T and Bob\'ak A 2011 \JMMM {\bf 323} 2095--102.
\bibitem{Fis60a}
Fisher M E 1960 {\it Proc. R. Soc. London} A {\bf 254} 66--85.
\bibitem{Fis60b}
Fisher M E 1960 {\it Proc. R. Soc. London} A {\bf 256} 502--13.
\bibitem{Hat68}
Hattori M and Nakano H 1968 {\it Progr. Theor. Phys.} {\bf 40} 958--74.
\bibitem{Mas73}
Mashiyama H and Nara S 1973 \PR B {\bf 7} 3119.
\bibitem{Lu05}
Lu W T and Wu F Y 2005 \PR E {\bf 71} 046120.
\bibitem{Can06}
\v{C}anov\'a L and Ja\v{s}\v{c}ur M 2006 {\it Condens. Matter Phys.} {\bf 9} 47--54.
\bibitem{Fis59}
Fisher M E 1959 \PR {\bf 113} 969.
\bibitem{Syo72}
Syozi I 1972 {\it Phase Transition and Critical Phenomena} (New York: Academic Press) pp. 269--329.
\bibitem{Str10}
Stre\v{c}ka J 2010 \PL A {\bf 374} 3718--22.
\bibitem{note1}
Explicit forms of the mapping parameters $A$, $J_{eff}$ are given by Eqs.~(7) and~(8) in Ref.~\cite{Can06}, by keeping $D=0$ and $H=B$.
\bibitem{Ons44}
Onsager L 1944 \PR {\bf 65} 117.
\bibitem{Bar88}
Barry J H, Khatun M and Tanaka T 1988 \PR B {\bf 37} 5193.
\bibitem{Kha90}
Khatun M, Barry J H and Tanaka T 1990 \PR B {\bf 42} 4398.
\bibitem{Bar91}
Barry J H, Tanaka T, Khatun M and M\'unera C H 1991 \PR B {\bf 44} 2595.
\bibitem{Bar95}
Barry J H and Khatun M 1995 \PR B {\bf 51} 5840.
\bibitem{Cal63}
Callen H B 1963 \PL {\bf 4} 161.
\bibitem{Suz65}
Suzuki M 1965 \PL {\bf 19} 267.
\bibitem{Bal02}
Balcerzak T 2002 \JMMM {\bf 246} 213--22.
\bibitem{Kau49}
Kaufman B and Onsager L 1949 \PR {\bf 76} 1244.
\bibitem{Yan52}
Yang C N 1952 \PR {\bf 85} 808.
\bibitem{note2}
Because of we consider the generalization of the original Fisher's super-exchange model in terms of the magnitude $s$ of decorating spins, the presented analysis also includes the particular spin case $s = 1/2$, which was discussed by M. E. Fisher several decades ago~\cite{Fis60a, Fis60b}.
\bibitem{Nee48}
N\'eel L 1948 {\it Ann. Phys. (Paris)} {\bf 3} 137--98.
\bibitem{Chi97}
Chikazumi S 1997 {\it Physics of Ferromagnetism} (Oxford: Oxford University Press) p. 147.
\bibitem{Str06}
Stre\v{c}ka J 2006 {\it Physica} A {\bf 360} 379--90.
\end{thebibliography}

\end{document}